\documentclass[aps,
twocolumn,
showpacs,pra,amssymb, amsmath]{revtex4}
\usepackage{amsmath,latexsym,amssymb}
\usepackage[dvips]{graphicx}
\usepackage{amsmath}
\usepackage{latexsym}
\usepackage{amssymb}
\usepackage{bm}

\begin{document}

\title{The conflict 
between Bell-\.Zukowski inequality and Bell-Mermin inequality}

\author{Koji Nagata}
\affiliation{ Department of Physics, Korea Advanced Institute of
Science and Technology, Daejeon 305-701, Korea}

\author{Jaewook Ahn}
\affiliation{ Department of Physics, Korea Advanced Institute of
Science and Technology, Daejeon 305-701, Korea}

\date{\today}

\begin{abstract}
We consider a two-particle/two-setting Bell experiment to visualize the
conflict between Bell-\.Zukowski inequality and Bell-Mermin inequality. 
The experiment
is reproducible by local realistic theories which are not 
rotationally invariant. We found that the
average value of the Bell-\.Zukowski operator can be evaluated only
by the two-particle/two-setting 
Bell experiment in question. The Bell-\.Zukowski
inequality reveals that the constructed local realistic models for the experiment are not 
rotationally invariant. That is, the 
two-particle Bell experiment in question reveals the conflict
between Bell-\.Zukowski inequality and Bell-Mermin inequality. Our analysis has found
the threshold visibility for the two-particle interference to reveal
the conflict noted above. It is found that the threshold visibility
agrees with the value to obtain a violation of the Bell-\.Zukowski
inequality.

\end{abstract}

\maketitle

\section{Introduction}
Local and realistic theories assume that physical properties exist
irrespective of whether they are measured and that the result of
measurement pertaining to one system is independent of any other
measurement simultaneously performed on a different system at a
distance. As Bell reported in 1964~\cite{bib:Bell}, certain
inequalities that correlation functions of a local realistic theory
must obey can be violated by quantum mechanics. Bell used the
singlet state to demonstrate this. Likewise, a
certain set of correlation functions produced by quantum
measurements of a single quantum state can contradict local
realistic theories. That is, we can see the conflict between
local realism and quantum mechanics. Since Bell work, local
realistic theories have been researched
extensively~\cite{bib:Redhead,bib:Peres}. Numerous experiments have
shown that Bell inequalities and local realistic theories are
violated~\cite{ex1,ex2,ex3,ex4,ex5,ex6,ex7,ex8,ex9}.

In 1982, Fine presented~\cite{Fine1,Fine2} the following example. A set
of correlation functions can be described with the property that
they are reproducible by local realistic theories for a system in
two-partite states if and only if the set of correlation functions
satisfies the complete set of (two-setting) Bell inequalities. This
is generalized to a system described by
multipartite~\cite{bib:Zukowski,gene1,gene2} states in the case where
two dichotomic observables are measured per site. We have,
therefore, obtained the necessary and sufficient condition for a set
of correlation functions to be reproducible by local realistic
theories in the specific case mentioned above.

A violation of `standard' two-settings 
Bell inequalities~\cite{bib:Zukowski,gene1,gene2,Mermin,Roy,Belinskii} 
is sufficient for experimentalists
to show the conflict between local realism and quantum mechanics.
However, it is necessary to create an entangled
state with sufficient visibility to violate a Bell inequality.
Furthermore, measurement settings should be established such that
the Bell inequality is violated. We consider, therefore, the
following question: What is a general method for experimentalists to
see the conflict between local realism and quantum mechanics only
from actually measured data?

In many cases one can build a local realistic model
for the observed data.
However, many such models are artificial
and can be disproved
if some principles
of physics are taken into account.
An example of such a principle
is rotational invariance of correlation function ---
the fact that the value of correlation function does not depend 
on the orientation of reference frames.
Taking this additional requirement into account
rules out local realistic models
even in situations
in which standard Bell inequalities allow for an explicit
construction of such models \cite{Nagata}.

In this paper, we study the physical phenomenon presented
in Ref.~\cite{Nagata}.
Here, we present a method using two Bell
operators~\cite{Bellope}. To this end, only a two-setting and
two-particle Bell experiment reproducible by local realistic
theories is needed. Such a Bell experiment also reveals, despite
appearances, the conflict between Bell-\.Zukowski inequality and 
Bell-Mermin inequality in the sense that the Bell-\.Zukowski inequality
\cite{bib:Zukowski2} is violated.

Our thesis is as follows. Consider two-qubit states that, under
specific settings, give correlation functions reproducible by specific local
realistic theory. Imagine that $N$ copies of the states can be
distributed among $2N$ parties, in such a way that each pair of
parties shares one copy of the state. The parties perform a
Bell-Greenberger-Horne-Zeilinger (GHZ) $2N$-particle
experiment~\cite{bib:Zukowski,gene1,gene2,Mermin,Roy,Belinskii} 
on their qubits.
Each of the pairs of parties uses the measurement settings noted
above. The Bell-Mermin operator, $B$, for their experiment does not
show violation of local realism. Nevertheless, one find another
Bell operator, which differs from $B$ by a numerical factor, that
does show such a violation.

This phenomenon can occur when the system is in a mixed two-qubit
state. We analyze threshold visibility for two-particle interference
to reveal the conflict mentioned above. It is found that the
threshold visibility agrees with the value to obtain a violation of
the Bell-\.Zukowski inequality.

\section{Experimental situation}

Consider two-qubit states:
\begin{eqnarray}
\rho_{a,b}=V|\psi\rangle\langle\psi|+(1-V)\rho_{\rm noise} ~
(0\leq V\leq 1),\label{qubit}
\end{eqnarray}
where $|\psi\rangle$ is Bell state as
$|\psi\rangle=\frac{1}{\sqrt{2}}(|+^a;+^b\rangle-i|-^a;-^b\rangle)$.
$\rho_{\rm noise} = \frac{1}{4} \openone$ is the random noise admixture. The value of $V$ can be interpreted as the reduction factor of the interferometric contrast observed in the two-particle correlation experiment.
The states $| \pm^k \rangle$ are eigenstates of $z$-component
Pauli observable $\sigma_z^k$ for the $k$th observer.
Here $a$ and $b$ are the label of parties (say Alice and Bob).
Then we have
${\rm tr}[\rho_{a,b}\sigma^a_x\sigma^b_x]=0$,
${\rm tr}[\rho_{a,b}\sigma^a_y\sigma^b_y]=0$,
${\rm tr}[\rho_{a,b}\sigma^a_x\sigma^b_y]=V$, and
${\rm tr}[\rho_{a,b}\sigma^a_y\sigma^b_x]=V$.
Here $\sigma^k_x$ and $\sigma^k_y$ are Pauli-spin operators for
$x$-component and for $y$-component,
respectively.
This set of experimental correlation functions is described
with the property that they are reproducible by two-settings local realistic theories.
See the following relations
along with the arguments in \cite{Fine1,Fine2}
\begin{widetext}
\begin{eqnarray}
&&|{\rm tr}[\rho_{a,b}\sigma^a_x\sigma^b_x]
-{\rm tr}[\rho_{a,b}\sigma^a_y\sigma^b_y]
+{\rm tr}[\rho_{a,b}\sigma^a_x\sigma^b_y]
+{\rm tr}[\rho_{a,b}\sigma^a_y\sigma^b_x]|=2V \leq 2,\nonumber\\
&&|{\rm tr}[\rho_{a,b}\sigma^a_x\sigma^b_x]
+{\rm tr}[\rho_{a,b}\sigma^a_y\sigma^b_y]
-{\rm tr}[\rho_{a,b}\sigma^a_x\sigma^b_y]
+{\rm tr}[\rho_{a,b}\sigma^a_y\sigma^b_x]|=0 \leq 2,\nonumber\\
&&|{\rm tr}[\rho_{a,b}\sigma^a_x\sigma^b_x]
+{\rm tr}[\rho_{a,b}\sigma^a_y\sigma^b_y]
+{\rm tr}[\rho_{a,b}\sigma^a_x\sigma^b_y]
-{\rm tr}[\rho_{a,b}\sigma^a_y\sigma^b_x]|=0\leq 2,\nonumber\\
&&|{\rm tr}[\rho_{a,b}\sigma^a_x\sigma^b_x]
-{\rm tr}[\rho_{a,b}\sigma^a_y\sigma^b_y]
-{\rm tr}[\rho_{a,b}\sigma^a_x\sigma^b_y]
-{\rm tr}[\rho_{a,b}\sigma^a_y\sigma^b_x]|=2V \leq 2.\label{LHV}
\end{eqnarray}
\end{widetext}

In the following section, we will use this kind of experimental
situation. Interestingly, those experimental correlation functions
can compute a violation of the Bell-\.Zukowski inequality. In order
to do so, we will present Bell operator method for such experimental
data to reveal that constructed local realistic models are not rotationally 
invariant if
$V>2(2/\pi)^2 \simeq 0.81$. Of course, such conflict between Bell-\.Zukowski inequality and Bell-Mermin inequality is derived only from actually measured
data which is modeled by two-settings local realistic theories.

\section{conflict between local realism and
quantum mechanics}

Let ${\bf N}_{2N}$ be $\{1,2,\ldots,2N\}$. Imagine that $N$ copies
of the states introduced in the preceding section can be distributed
among $2N$ parties, in such a way that each pair of parties shares
one copy of the state
\begin{eqnarray}
\rho^{\otimes N}=\underbrace{\rho_{1,2}\otimes\rho_{3,4}\otimes
\cdots\otimes\rho_{N-1,N}}_{N}.
\end{eqnarray}
Suppose that spatially separated $2N$ observers perform measurements
on each of $2N$ particles. The decision processes for choosing
measurement observables are spacelike separated.

We assume that a two-orthogonal-setting Bell-GHZ $2N$-particle
correlation experiment is
performed. We choose measurement observables such that
\begin{eqnarray}
A_k=\sigma^k_x,
A'_k=\sigma^k_y.
\label{setting}
\end{eqnarray}
Namely, each of the pairs of parties uses measurement settings such
that they can check the condition (\ref{LHV}). Therefore, it should
be that given $2^{2N}$ correlation functions are described with the
property that they are reproducible by two-settings local realistic theories.

Bell-Mermin operators ${B}_{{\bf N}_{2N}}$ and ${B}'_{{\bf N}_{2N}}$
(defined as follows) do not show any violation of local realism as
shown below.

Let $f(x,y)$ denote the function
$\frac{1}{\sqrt{2}}e^{-i\pi/4}(x+iy), x,y\in {\bf R}$. $f(x,y)$ is
invertible as $ x=\Re f-\Im f, y=\Re f+\Im f $. Bell-Mermin
operators ${B}_{{\bf N}_{2N}}$ and ${B}'_{{\bf N}_{2N}}$ are defined
by \cite{Mermin,Roy,Belinskii,bib:Werner2} $ f({B}_{{\bf N}_{2N}},{B}'_{{\bf
N}_{2N}})=\otimes_{k=1}^{2N} f(A_k,A'_k) $. Bell-Mermin inequality
can be expressed as \cite{bib:Werner2}
\begin{eqnarray}
|\langle{B}_{{\bf N}_{2N}}\rangle|\leq 1, ~~|\langle{B}'_{{\bf N}_{2N}}
\rangle|\leq 1,\label{MKin}
\end{eqnarray}
where ${B}_{{\bf N}_{2N}}$ and ${B}'_{{\bf N}_{2N}}$ are Bell-Mermin
operators defined by
\begin{eqnarray}
f({B}_{{\bf N}_{2N}},{B}'_{{\bf N}_{2N}})=\otimes_{k=1}^{2N}
f(A_k,A'_k).\label{MK}
\end{eqnarray}
We also define ${B}_{\alpha}$ for any subset $\alpha\subset{\bf N}_{2N}$ by
\begin{eqnarray}
f({B}_{\alpha},{B}'_{\alpha})=\otimes_{k\in \alpha}
f(A_k,A'_k).
\end{eqnarray}
It is easy to see that, when $\alpha, \beta(\subset{\bf N}_{2N})$ are disjoint,
\begin{eqnarray}
f({B}_{\alpha\cup \beta},
{B}'_{\alpha\cup \beta})=f({B}_{\alpha},{B}'_{\alpha})\otimes
f({B}_{\beta},
{B}'_{\beta}),
\end{eqnarray}
which leads to following equations,
\begin{eqnarray}
{B}_{\alpha\cup \beta}&=&
(1/2){B}_{\alpha}\otimes({B}_{\beta}
+{B}'_{\beta})
+
(1/2){B}'_{\alpha}\otimes({B}_{\beta}-
{B}'_{\beta}),\nonumber\\
{B}'_{\alpha\cup \beta}&=&
(1/2){B}'_{\alpha}\otimes({B}'_{\beta}
+{B}_{\beta})
+
(1/2){B}_{\alpha}\otimes({B}'_{\beta}-
{B}_{\beta}).\nonumber\\
\label{GB}
\end{eqnarray}
In specific operators $A_k,A'_k$ given in Eq.~(\ref{setting}),
where $\sigma^k_x=|+^k\rangle\langle-^k|+|-^k\rangle\langle+^k|$ and
$\sigma^k_y=-i|+^k\rangle\langle-^k|+i|-^k\rangle\langle+^k|$, we
have (cf. \cite{bib:scarani})
\begin{eqnarray}
&&f(A_k,A'_k)=(e^{-i\frac{\pi}{4}}/\sqrt{2})
(\sigma_x^k+i\sigma^k_y)\nonumber\\
&&=e^{-i\frac{\pi}{4}}\sqrt{2}|+^k\rangle\langle-^k|
\end{eqnarray}
and
\begin{eqnarray}
&&f({B}_{{\bf N}_{2N}},{B}'_{{\bf N}_{2N}})=
\otimes_{k=1}^{2N} f(A_k,A'_k)\nonumber\\
&&=e^{-i\frac{2N\pi}{4}} 2^{N}\otimes_{k=1}^{2N} |+^k\rangle\langle-^k|
\nonumber\\
&&=e^{-i\frac{2N\pi}{4}} 2^{N}|+^{\otimes 2N}\rangle\langle-^{\otimes 2N}|.
\end{eqnarray}
Hence we obtain
\begin{eqnarray}
{B}_{{\bf N}_{2N}}
=2^{(2N-1)/2}(|\Psi^{+}_0\rangle\langle \Psi^{+}_0|-
|\Psi^{-}_0\rangle\langle \Psi^{-}_0|),\label{BMineq}
\end{eqnarray}
where $e^{-i\frac{(2N-1)\pi}{4}}|+^{\otimes 2N}\rangle=
|1^{\otimes 2N}\rangle$.
Here the states $|\Psi^{\pm}_0\rangle$ are Greenberger-Horne-Zeilinger (GHZ)
states \cite{bib:GHZ}, i.e.,
\begin{eqnarray}
|\Psi^{\pm}_0\rangle=\frac{1}{\sqrt{2}}
(|0^{\otimes 2N}\rangle\pm|1^{\otimes 2N}\rangle).
\end{eqnarray}

Measurements on each of $2N$ particles enable them
to obtain $2^{2N}$ correlation functions.
Thus, they get an average value of
specific Bell-Mermin operator given in Eq.~(\ref{BMineq}).
According to Eq.~(\ref{GB}), we obtain
\begin{eqnarray}
\langle {B}_{{\bf N}_{2N}} \rangle=
\langle {B}'_{{\bf N}_{2N}} \rangle=
\prod_{i=2}^{N}\langle {B}_{\{i-1,i\}} \rangle=
V^{N}(\leq 1).
\end{eqnarray}
Clearly, Bell-Mermin operators, ${B}_{{\bf N}_{2N}}$
and ${B}'_{{\bf N}_{2N}}$,
for their experiment do not show any violation of local realism
as we have mentioned above.

Nevertheless, one can find $2N$-partite Bell operator, which one may
call Bell-\.Zukowski operator $Z_{2N}$, which differs from
${B}_{{\bf N}_{2N}}$ only by a numerical factor, that does show such
a violation. Bell-\.Zukowski operator ${Z}_{2N}$ is as (cf. Appendix
\ref{three}, Eq.~(\ref{final2}))
\begin{eqnarray}
{Z}_{2N}=\frac{1}{2}\left(\frac{\pi}{2}\right)^{2N}
(|\Psi^+_0\rangle\langle\Psi^+_0|-
|\Psi^-_0\rangle\langle\Psi^-_0|).\label{ZukoBellope}
\end{eqnarray}
Clearly,
we see that Bell-Mermin operator given in Eq.~(\ref{BMineq})
is connected to Bell-\.Zukowski operator
${Z}_{2N}$ in the following relation
\begin{eqnarray}
&&{Z}_{2N}=\frac{1}{2}\left(\frac{\pi}{2}\right)^{2N}
\frac{1}{2^{(2N-1)/2}}
 {B}_{{\bf N}_{2N}}.\label{Bellrelation}
\end{eqnarray}
One can see that specific two settings Bell $2N$-particle experiment
in question computes an average value of Bell-\.Zukowski operator
$\langle{Z}_{2N}\rangle$ via an average value of $\langle {B}_{{\bf
N}_{2N}}\rangle$.

Therefore, from the
Bell-\.Zukowski inequality $|\langle {Z}_{2N}\rangle|\leq 1$,
we have a condition on the average value of Bell-Mermin operator $\langle {B}_{{\bf N}_{2N}}\rangle$ which is written by
\begin{eqnarray}
|\langle {B}_{{\bf N}_{2N}}\rangle|\leq
2\left(\frac{2}{\pi}\right)^{2N} 2^{({2N}-1)/2}.\label{newbell}
\end{eqnarray}
Please notice that the Bell-\.Zukowski inequality $|\langle
Z_{2N}\rangle|\leq 1$ is derived under the assumption that there are
predetermined `hidden' results of the measurement for all directions
in the rotation plane for the system in a state. On the other hand,
Bell-Mermin inequality is derived under the assumption that there
are predetermined `hidden' results of the measurement for two
directions for the system in a state.
Thus, Bell-\.Zukowski inequality governs rotationally invariant
descriptions while Bell-Mermin inequality does not.

When $N\geq 2$ and $V$ is given by
\begin{equation}
\left(2\left(\frac{2}{\pi}\right)^{2N} 2^{(2N-1)/2}\right)^{1/N}<V \leq 1,
\label{mainresult}
\end{equation}
one can {\it compute} a violation of the Bell-\.Zukowski inequality
$|\langle {Z}_{2N}\rangle|\leq 1$, that is, 
the explicit local realistic models 
are not rotationally invariant. The condition
(\ref{mainresult}) says that threshold visibility decreases when the
number of copies $N$ increases. In extreme situation, when
$N\rightarrow \infty$, we have desired condition $V>2(2/\pi)^2$ to
show the conflict in question. It agrees with the value to get a
violation of the Bell-\.Zukowski inequality.
(It also agrees with the value to get a
violation of the generalized Bell inequality presented in 
Ref.~\cite{Nagata}.)

Thus the given example using two-qubit states reveals the violation
of the Bell-\.Zukowski inequality. The interesting point is that all
the information to get the violation of the Bell-\.Zukowski
inequality can be obtained only by a two-setting and two-particle
Bell experiment reproducible by two-settings local realistic theories.

It presents a quantum-state measurement
situation that admits local realist descriptions for the given apparatus
settings, but no local realist descriptions which are rotationally
invariant, even though the experiment should be ruled by 
rotationally
invariant laws.  There is no local realist theory for the 
experiment as a
whole and so such a descriptions is only possible for 
certain setting.

What the result illustrates is
that there is a further division among the measurement 
settings, those
that admit rotational invariant local realist models and 
those that do
not. This is another manifestation of the underlying 
contextual nature
of realist theories of quantum experiments.


\section{summary }

In summary, we have presented a Bell operator method. This approach
provides a means to check if the explicit model is rotationally invariant, 
i.e., if a conflict between Bell-\.Zukowski inequality and Bell-Mermin inequality occurs. Our argument relies only on a
two-setting and two-particle Bell experiment reproducible by a local
realistic theory which is not rotationally invariant. 
Given a two-setting and two-particle Bell
experiment reproducible by specific local realistic theory, one can compute
a violation of Bell-\.Zukowkski inequality. Measured data indicates
that the explicit local realistic models are not rotationally invariant.
Thus, the conflict between Bell-\.Zukowski inequality and Bell-Mermin inequality is,
despite appearances, revealed.

This phenomenon can occur when the system is in a mixed state. We
also analyzed the threshold visibility for two-particle interference
in order to bring about the phenomenon. The threshold visibility
agrees well with the value to obtain a violation of the
Bell-\.Zukowski inequality.


\section*{Acknowledgments}
This work has been supported by Frontier Basic Research Programs at
KAIST and K.N. is supported by a BK21 research grant.

\appendix

\section{Bell-\.Zukowski inequality}
\label{three}

Let us review the Bell-\.Zukowski inequality proposed in
Ref.~\cite{bib:Zukowski2}. Let ${L}({H})$ be the space of Hermitian
operators acting on a finite-dimensional Hilbert space ${H}$, and
${T}({H})$ be the space of density operators acting on the Hilbert
space ${H}$. Namely, ${T}({H})= \{\rho |
\rho\in{L}({H})\wedge\rho\geq 0\wedge {\rm tr}[\rho]=1\}$. We also
consider a classical probability space $(\Omega,\Sigma,M_{\rho})$,
where $\Omega$ is a nonempty space, $\Sigma$ is a $\sigma$-algebra
of subsets of $\Omega$, and $M_{\rho}$ is a $\sigma$-additive
normalized measure on $\Sigma$ such that $M_{\rho}(\Omega)=1$. The
subscript $\rho$ expresses following meaning. The probability
measure $M_{\rho}$ is determined uniquely when a state $\rho$ is
specified.

Consider a quantum state $\rho$ in ${T}(\otimes_{k=1}^n{H}_k)$,
where ${H}_k$ represents the Hilbert space with respect to party
$k\in{\bf N}_n(=\{1,2,\ldots,n\})$. Then we can define measurable
functions $f_k: o_k,\omega \mapsto f_k(o_k,\omega)\in
{[I(o_k),S(o_k)]}, o_k\in{L}({H}_k), \omega\in \Omega$. Here
$S(o_k)$ and $I(o_k)$ are the supremum and the infimum of the
spectrum of $o_k\in{L}({H}_k)$, respectively. Those functions
$f_k(o_k,\omega)$ must not depend on the choices of $v$'s on the
other sites in ${\bf N}_n\backslash\{k\}$. Using the functions
$f_k$, we define a quantum correlation function which admits a local
realistic model \cite{WERNER1}.

{\it Definition 1}. A quantum correlation function ${\rm
tr}[\rho\otimes_{k=1}^n o_k]$ is said to admit a local realistic
model if and only if there exist a classical probability space
$(\Omega,\Sigma,M_{\rho})$ and a set of functions
$f_1,f_2,\ldots,f_n$, such that
\begin{eqnarray}
\int_{\Omega}\!\! M_{\rho}(d\omega)\prod_{k=1}^n
f_k(o_k,\omega)
={\rm tr}[\rho \otimes_{k=1}^n o_k]\label{definition2}
\end{eqnarray}
for a Hermitian operator $\otimes_{k=1}^n o_k$, where $o_k \in
{L}({H}_k)$. Note that there are several (noncommuting) observables
per site, however above definition is available for just one $o_k$
per site.

We consider a situation where each of the $n$ spatially separated
observers has infinite number of settings of measurements (in the
$xy$ plane) to choose from. The operation of each of the measuring
apparatuses is controlled by a knob. The knob sets a parameter
$\phi$. An apparatus performs measurements of a Hermitian operator
$\sigma_{\phi}$ on two-dimensional space with two eigenvalues $\pm
1$. The corresponding eigenstates are defined as $
|\pm;\phi\rangle=(1/\sqrt{2})(|1\rangle\pm e^{(i\phi)}|0\rangle). $
The local phases that they are allowed to set are chosen as $ 0\leq
\phi^k< \pi $ for the $k$th observer. The Bell-\.Zukowski inequality
can be written as
\begin{eqnarray}
|\langle {Z}_n\rangle|\leq 1,\label{ZukowskiBell2}
\end{eqnarray}
where the corresponding Bell
operator ${Z}_n$ is
\begin{eqnarray}
{Z}_n
=\left(\frac{1}{2^{n}}\right)
\int_0^{\pi}\!\!\!\!d\phi^1\cdots\int_0^{\pi}\!\!\!\!d\phi^n
\cos\left(\sum_{k=1}^n \phi^k\right)
\otimes_{k=1}^n\sigma_{\phi^k},\nonumber\\
\label{Zukowskiope2}
\end{eqnarray}
where
\begin{eqnarray}
&&\sigma_{\phi^k}=e^{-i\phi^k}|1^k\rangle\langle 0^k|+
e^{i\phi^k}|0^k\rangle\langle 1^k|,
 k\in {\bf N}_n.\label{Pauliope2}
\end{eqnarray}
Bell-\.Zukowski operator ${Z}_n$ is a sum of
infinite number of Hermitian operators,
except for fixed number $1/(2^{n})$.
We shall mention why ${Z}_n$ given in Eq.~(\ref{Zukowskiope2})
is a Bell operator
when Eq.~(\ref{ZukowskiBell2}) is a Bell inequality as follows.

Let us assume that all of quantum correlation functions (every
setting lies in $xy$ plane) admit a local realistic model. Here each
party $k$ performs locally measurements on an arbitrary single state
$\rho$.

Then, according to {\it Definition 1} (Eq.~(\ref{definition2})),
there exists a classical probability space
$(\Omega,\Sigma,M_{\rho})$ related to the state in question $\rho$.
And there exists a set of functions $f_1,f_2,\ldots,f_n(\in[-1,1])$
such that
\begin{eqnarray}
\int_{\Omega}\!\! M_{\rho}(d\omega)\prod_{k=1}^n
f_k(\sigma_{\phi^k},\omega)
={\rm tr}[\rho \otimes_{k=1}^n\sigma_{\phi^k}]\label{integral2}
\end{eqnarray}
for every $0\leq \phi^k< \pi,~  k\in{\bf N}_n$.
Hence an expectation of
a sum of infinite number of Hermitian operators
(i.e., $2^{n}{Z}_n$) is bounded by the
possible values of
\begin{eqnarray}
S^{(\infty, n)}_{\omega}&=&\int_0^{\pi}d\phi^1\cdots\int_0^{\pi}d\phi^n
\left[
\cos\left(\sum_{k=1}^n \phi^k\right)
\prod^n_{k=1}f_k(\sigma_{\phi^k},\omega)\right]\nonumber\\
&=&\Re \left(\prod_{k=1}^n z'_k\right),
\end{eqnarray}
where $z'_k=\int_0^{\pi}d\phi^k f_k(\sigma_{\phi^k},\omega)
\exp\left(i \phi^k\right)$.

Let us derive
an upper bound of $S^{(\infty, n)}_{\omega}$.
We may
assume $f_k =\pm 1$.
Let us analyze the structure of the following integral
\begin{eqnarray}
&&z'_k=\int_0^{\pi}\!\!\!\!d\phi^k f_k(\sigma_{\phi^k},\omega)
\exp\left(i \phi^k\right)\nonumber\\
&&=\int_0^{\pi}\!\!\!\!d\phi^k f_k(\sigma_{\phi^k},\omega)
(\cos\phi^k+i\sin\phi^k).\label{integral}
\end{eqnarray}
Notice that Eq.~(\ref{integral}) is a sum of the following integrals:
\begin{eqnarray}
\int_0^{\pi}\!\!\!\!d\phi^k f_k(\phi^k, \omega) \cos \phi^k
\end{eqnarray}
and
\begin{eqnarray}
\int_0^{\pi}\!\!\!\!d\phi^k f_k(\phi^k, \omega) \sin \phi^k.
\end{eqnarray}
We deal here with integrals, or rather scalar products of
$f_k(\phi^k, \omega) $ with two orthogonal functions.
One has
\begin{eqnarray}
\int_0^{\pi}\!\!\!\!d\phi^k \cos \phi^k\sin \phi^k=0.
\end{eqnarray}
The normalized functions $\frac{1}{\sqrt{\pi/2}}\cos\phi^k$ and
$\frac{1}{\sqrt{\pi/2}}\sin\phi^k$ form a basis of a real
two-dimensional functional space, which we shall call $S^{(2)}$.
Note further that any function in $S^{(2)}$ is of the form
\begin{eqnarray}
A\frac{1}{\sqrt{\pi/2}}\cos\phi^k +
B\frac{1}{\sqrt{\pi/2}}\sin\phi^k,
\end{eqnarray}
where $A$ and $B$ are constants,
and that any normalized function in $S^{(2)}$ is given by
\begin{eqnarray}
&&\cos \psi \frac{1}{\sqrt{\pi/2}}\cos\phi^k
+ \sin \psi \frac{1}{\sqrt{\pi/2}}\sin\phi^k\nonumber\\
&&=\frac{1}{\sqrt{\pi/2}}\cos(\phi^k-\psi).
\end{eqnarray}
The norm $\Vert f_{k}^{||} \Vert$ of the projection of $f_k$ into
the space $S^{(2)}$ is given by the maximal possible value of the
scalar product $f_k$ with any normalized function belonging to
$S^{(2)}$, that is
\begin{eqnarray}
\Vert f_{k}^{||} \Vert= \max_{\psi}
\int_0^{\pi}\!\!\!\!d\phi^k
f_k(\phi^k, \omega)
\frac{1}{\sqrt{\pi/2}}\cos (\phi^k-\psi).
\end{eqnarray}
Because $|f_k(\phi^k, \omega)|=1$,
one has $\Vert f_{k}^{||} \Vert\leq 2/\sqrt{\pi/2}$.
Since $\frac{1}{\sqrt{\pi/2}}\cos\phi^k$
and $\frac{1}{\sqrt{\pi/2}}\sin\phi^k$
are two orthogonal basis functions in $S^{(2)}$, one has
\begin{eqnarray}
\int_0^{\pi}\!\!\!\!d\phi^k
f_k(\phi^k, \omega)
\frac{1}{\sqrt{\pi/2}}\cos \phi^k
=\cos \beta_k \Vert f_{k}^{||} \Vert
\end{eqnarray}
and
\begin{eqnarray}
\int_0^{\pi}\!\!\!\!d\phi^k
f_k(\phi^k, \omega)
\frac{1}{\sqrt{\pi/2}}\sin \phi^k=\sin \beta_k \Vert f_{k}^{||} \Vert,
\end{eqnarray}
where $\beta_k$ is some angle. Using this fact, one can put the
value of (\ref{integral}) into the following form
\begin{eqnarray}
&&z'_k=\sqrt{\pi/2}
\Vert f_{k}^{||} \Vert(\cos\beta_k+i\sin\beta_k)\nonumber\\
&&=\sqrt{\pi/2}
\Vert f_{k}^{||} \Vert\exp\left(i \beta_k\right).
\end{eqnarray}
Therefore, since $\Vert f_{k}^{||} \Vert\leq 2/\sqrt{\pi/2}$, the maximal value
of $|z'_k|$ is 2.
Hence, we have $|\prod_{k=1}^n z'_k|\leq 2^n$.
Then we get
\begin{eqnarray}
|S^{(\infty, n)}_{\omega}|\leq 2^n.\label{relation3}
\end{eqnarray}
Let $E(\cdot)$ represent an expectation on the classical probability
space. If we integrate this relation (\ref{relation3}) under
 normalized measure $M_{\rho}(d\omega)$ over a space $\Omega$,
we obtain the relation (\ref{ZukowskiBell2}). Here we have used the
relation that $E(S^{(\infty, n)}_{\omega})=2^{n}{\rm tr}[\rho{Z}_n]$
(see Eq.~(\ref{integral2})). Therefore, we have proven the
Bell-\.Zukowski inequality (\ref{ZukowskiBell2}) from an assumption.
The assumption is that all of infinite number of quantum correlation
functions (every setting lies in $xy$ plane) admit a local realistic
model.

Let us consider matrix elements of Bell-\.Zukowski operator ${Z}_n$
as given in Eq.~(\ref{Zukowskiope2}) on using GHZ basis
\begin{eqnarray}
|\Psi_j^{\pm}\rangle=
\frac{1}{\sqrt{2}}(|j\rangle|0\rangle\pm
|2^{n-1}-j-1\rangle|1\rangle),\label{GHZbasis}
\end{eqnarray}
where $j=j_1 j_2 \cdots j_{n-1}$ is understood in binary notation.
It is clear that no off-diagonal element appears, because of the
form of the operator $\sigma_{\phi^k}$ as given in
Eq.~(\ref{Pauliope2}).

Let $\beta$ be a subset $\beta\subset{\bf N}_n$ and $l(\beta)$ be an
integer $l_1\cdots l_n$ in the binary notation with $l_m=1$ for
$m\in \beta$ and $l_m=0$ otherwise. And let $j(\beta)$ be an integer
binary-represented by $l_1\cdots l_{n-1}$. Then we define a
two-to-one function $g:\beta\mapsto g(\beta)\in \{0\} \cup{\bf
N}_{2^{(n-1)}-1}$ where $g(\beta)$ takes the values $j(\beta)$ and
$2^{n-1}-j(\beta)-1$, respectively, for even and odd values of
$l(\beta)$.

In what follows, we show that
$\langle\Phi^{\pm}_{g(\alpha)}|{Z}_n|\Phi^{\pm}_{g(\alpha)}\rangle=0$
 for any subset
$\alpha\subset{\bf N}_n$ when $\alpha\neq \emptyset, {\bf N}_n$.
We also show that $\langle\Psi^{\pm}_{g(\alpha)}|{Z}_n
|\Psi^{\pm}_{g(\alpha)}\rangle= \pm
\frac{1}{\sqrt{2}}\left(\frac{\pi}{2}\right)^n$ when
$\alpha=\emptyset$ or $\alpha={\bf N}_n$. When $\alpha=\emptyset$ or
$\alpha={\bf N}_n$, we have
\begin{eqnarray}
&&2^{n}
\langle\Psi^{\pm}_0|{Z}_n|\Psi^{\pm}_0\rangle\nonumber\\
&=&\pm\int_0^{\pi}\!\!\!\!d\phi^1\cdots\int_0^{\pi}\!\!\!\!d\phi^n
\cos^2\left(\sum_{k=1}^n \phi^k\right)
\nonumber\\
&=&\pm\frac{1}{2}\int_0^{\pi}\!\!\!\!d\phi^1\cdots\int_0^{\pi}\!\!\!\!d\phi^n
\left[1+
\cos\left(2\sum_{k=1}^n \phi^k\right)\right]
\nonumber\\
&=&\pm\frac{1}{2}
\Re \left\{\int_0^{\pi}\!\!\!\!d\phi^1\cdots\int_0^{\pi}\!\!\!\!d\phi^n
\left[1+
\exp\left(2i\sum_{k=1}^n \phi^k\right)\right]\right\}
\nonumber\\
&=&\pm\frac{\pi^n}{2}\pm
\frac{1}{2}\Re \left(\prod_{k=1}^n\int_0^{\pi}\!\!\!\!d\phi^k
\exp\left(2i \phi^k\right)\right).\nonumber\\
\end{eqnarray}
Since $\int_0^{\pi}d\phi^k
\exp\left(2i \phi^k\right)=0, k\in{\bf N}_n$,
the last term vanishes.
Hence we get
\begin{eqnarray}
\langle\Psi^{\pm}_0|{Z}_n|\Psi^{\pm}_0\rangle
=\pm\frac{1}{2}\left(\frac{\pi}{2}\right)^n.
\end{eqnarray}
On the other hand, when $\alpha\neq \emptyset, {\bf N}_n$, we obtain
\begin{widetext}
\begin{eqnarray}
2^{n}
|\langle\Phi^{\pm}_{g(\alpha)}|{Z}_n|\Phi^{\pm}_{g(\alpha)}\rangle|
&=&\int_0^{\pi}\!\!\!\!d\phi^1\cdots\int_0^{\pi}\!\!\!\!d\phi^n
\cos\left(\sum_{k\in\alpha} \phi^k+ \sum_{k\in{\bf
N}_{n}\backslash{\alpha}} \phi^k\right)
\times\cos\left(\sum_{k\in\alpha} \phi^k- \sum_{k\in{\bf
N}_{n}\backslash{\alpha}} \phi^k\right)
\nonumber\\
&=&\frac{1}{2}\int_0^{\pi}\!\!\!\!d\phi^1\cdots\int_0^{\pi}\!\!\!\!d\phi^n
\left[
\cos\left(2\sum_{k\in\alpha} \phi^k\right)+
\cos\left(2\sum_{k\in{\bf N}_{n}\backslash{\alpha}}
\phi^k\right)\right]
\nonumber\\
&=&\frac{\pi^{|{\bf N}_{n}\backslash{\alpha}|}}{2}
\Re \left(\prod_{k\in\alpha}\int_0^{\pi}\!\!\!\!d\phi^k
\exp\left(2i \phi^k\right)\right)
+\frac{\pi^{|{\alpha}|}}{2}\Re \left(\prod_{k\in{\bf N}_{n}\backslash{\alpha}}
\int_0^{\pi}\!\!\!\!d\phi^k
\exp\left(2i \phi^k\right)\right).\nonumber\\
\end{eqnarray}
\end{widetext}
Since $\int_0^{\pi}d\phi^k
\exp\left(2i \phi^k\right)=0, k\in{\bf N}_n$,
the last two terms vanish.

Hence, Bell operator ${Z}_n$
as given in Eq.~(\ref{Zukowskiope2})
 can be rewritten as
\begin{eqnarray}
{Z}_n=\frac{1}{2}\left(\frac{\pi}{2}\right)^n
(|\Psi^+_0\rangle\langle\Psi^+_0|-
|\Psi^-_0\rangle\langle\Psi^-_0|).\label{final2}
\end{eqnarray}

\end{document}